\documentstyle[epsfig]{mn}

\newif\ifAMStwofonts

\def\lesssim{\mathrel{\hbox{\rlap{\hbox{\lower4pt\hbox{$\sim$}}}\hbox{$<$}}}}
\def\gtrsim{\mathrel{\hbox{\rlap{\hbox{\lower4pt\hbox{$\sim$}}}\hbox{$>$}}}}

\ifoldfss
  \ifCUPmtlplainloaded \else
    \NewTextAlphabet{textbfit} {cmbxti10} {}
    \NewTextAlphabet{textbfss} {cmssbx10} {}
    \NewMathAlphabet{mathbfit} {cmbxti10} {} 
    \NewMathAlphabet{mathbfss} {cmssbx10} {} 
  \fi
  \ifAMStwofonts
    \ifCUPmtlplainloaded \else
      \NewSymbolFont{upmath} {eurm10}
      \NewSymbolFont{AMSa} {msam10}
      \NewMathSymbol{\upi}     {0}{upmath}{19}
      \NewMathSymbol{\umu}     {0}{upmath}{16}
      \NewMathSymbol{\upartial}{0}{upmath}{40}
      \NewMathSymbol{\leqslant}{3}{AMSa}{36}
      \NewMathSymbol{\geqslant}{3}{AMSa}{3E}

       \let\le=\leqslant
       
    \fi
  \fi
\fi 

\ifnfssone
  \newmathalphabet{\mathit}
  \addtoversion{normal}{\mathit}{cmr}{m}{it}
  \addtoversion{bold}{\mathit}{cmr}{bx}{it}
  \newmathalphabet{\mathbfit} 
  \addtoversion{normal}{\mathbfit}{cmr}{bx}{it}
  \addtoversion{bold}{\mathbfit}{cmr}{bx}{it}
  \newmathalphabet{\mathbfss} 
  \addtoversion{normal}{\mathbfss}{cmss}{bx}{n}
  \addtoversion{bold}{\mathbfss}{cmss}{bx}{n}
  \ifAMStwofonts
    \ifCUPmtlplainloaded \else
      \UseAMStwoboldmath
      \makeatletter
      \new@mathgroup\upmath@group
      \define@mathgroup\mv@normal\upmath@group{eur}{m}{n}
      \define@mathgroup\mv@bold\upmath@group{eur}{b}{n}
      \edef\UPM{\hexnumber\upmath@group}
      \new@mathgroup\amsa@group
      \define@mathgroup\mv@normal\amsa@group{msa}{m}{n}
      \define@mathgroup\mv@bold\amsa@group{msa}{m}{n}
      \edef\AMSa{\hexnumber\amsa@group}
      \makeatother
      \mathchardef\upi="0\UPM19
      \mathchardef\umu="0\UPM16
      \mathchardef\upartial="0\UPM40
      \mathchardef\leqslant="3\AMSa36
      \mathchardef\geqslant="3\AMSa3E

       \let\le=\leqslant

    \fi
  \fi
\fi 

\ifnfsstwo
  \DeclareMathAlphabet{\mathbfit}{OT1}{cmr}{bx}{it}
  \SetMathAlphabet\mathbfit{bold}{OT1}{cmr}{bx}{it}
  \DeclareMathAlphabet{\mathbfss}{OT1}{cmss}{bx}{n}
  \SetMathAlphabet\mathbfss{bold}{OT1}{cmss}{bx}{n}
  \ifAMStwofonts
    \ifCUPmtlplainloaded \else
      \DeclareSymbolFont{UPM}{U}{eur}{m}{n}
      \SetSymbolFont{UPM}{bold}{U}{eur}{b}{n}
      \DeclareSymbolFont{AMSa}{U}{msa}{m}{n}
      \DeclareMathSymbol{\upi}{0}{UPM}{"19}
      \DeclareMathSymbol{\umu}{0}{UPM}{"16}
      \DeclareMathSymbol{\upartial}{0}{UPM}{"40}
      \DeclareMathSymbol{\leqslant}{3}{AMSa}{"36}
      \DeclareMathSymbol{\geqslant}{3}{AMSa}{"3E}

       \let\le=\leqslant

    \fi
  \fi
\fi 

\ifCUPmtlplainloaded \else
  \ifAMStwofonts \else 
    \def\upi{\pi}
    \def\umu{\mu}
    \def\upartial{\partial}
  \fi
\fi

\title[Probing Cosmic Reionization and Environments of GRBs through Radio Dispersion]
{Probing the Cosmic Reionization History
and Local Environment of Gamma-Ray Bursts
through Radio Dispersion}
\author[Susumu Inoue]
       {Susumu Inoue \thanks{inouemu@MPA-Garching.MPG.DE}\\
         Max-Planck-Institut f\"ur Astrophysik,
         Karl-Schwarzschild-Str. 1, Postfach 1317,
         85741 Garching, Germany}
\date{Accepted by MNRAS, 29 Oct 2003}

\pagerange{\pageref{firstpage}--\pageref{lastpage}}
\pubyear{2003}

\begin{document}

\maketitle

\label{firstpage}

\begin{abstract}
We discuss the effect of dispersion delay
due to intervening ionized media in the radio emission of gamma-ray bursts (GRBs).
For high redshift GRBs ($z \ga 3$), the ionized intergalactic medium (IGM)
should be the dominant source of dispersion without substantial local or foreground contamination,
offering a unique probe of the cosmic reionization epoch
through measures of the free electron column density out to different redshifts.
The delay times from $z \sim 10$ can be $\sim$ 1 hour at 100 MHz and $\sim$ 10 hours at 30 MHz.
On the other hand, dispersion by local ionized material may be important for GRBs at lower redshifts
if they occur inside or behind dense molecular clouds, providing clues to the GRB environment;
free-free absorption may also be significant in this case.
Detecting dispersion delay in the known radio afterglow emission should be extremely challenging
due to the low fluxes at the relevant frequencies,
but may be marginally possible for rare, bright afterglows
by the {\it Square Kilometer Array}.
If GRBs also emit prompt, coherent radio emission as predicted by Sagiv \& Waxman (2002),
the observational prospects can be much better;
detection of a sufficient sample of high $z$ GRBs by the {\it Low Frequency Array} 
may allow a discrimination of different reionization histories.
Interesting constraints on the low redshift, warm-hot IGM may also be obtainable through dispersion.
\end{abstract}


\section{Introduction}
\label{sec:intro}

Gamma-ray bursts (GRBs) are being increasingly recognized
as potentially powerful probes of the very high redshift universe
(Loeb 2003, Djorgovski et al. 2003).
The mounting evidence that GRBs (at least with long durations) are closely associated
with the formation and collapse of massive stars (e.g. van Paradijs, Kouveliotou \& Wijers 2000,
Bloom, Kulkarni \& Djorgovski 2002, Hjorth et al. 2003, Matheson et al. 2003)
indicates that they can occur in significant numbers at early epochs,
perhaps up to the formation era of the very first stars
(Lamb \& Reichart 2000, Bromm \& Loeb 2002).
Their bright, broadband emission, a large part of which can be robustly modeled
as nonthermal emission from a relativistic blast wave,
should be readily detectable out to the highest redshifts they are expected to exist
(Lamb \& Reichart 2000, Ciardi \& Loeb 2000).
Given the likelihood that bright quasars are much rarer at high $z$,
GRBs may play unique roles as beacons
illuminating the so-called ``dark ages'' of the universe
(Barkana \& Loeb 2001, Miralda-Escud\'e 2003).

An issue of considerable current interest is the epoch of cosmic reionization.
The question of when and how reionization in the universe occurred and what caused it is critical,
having profound implications for the formation of the first stars and galaxies and their subsequent evolution
(Madau \& Kuhlen 2003, Haiman 2003).
The highest redshift quasars discovered by the {\it Sloan Digital Sky Survey (SDSS)}
have revealed complete spectral absorption troughs (Gunn \& Peterson 1965),
showing that the neutral fraction of the intergalactic medium (IGM) is at least $\sim 10^{-3}$ at $z \sim 6$
(Djorgovski et al. 2001, Becker et al. 2001, White et al. 2003).
On the other hand, measurements of temperature-polarization correlations 
in cosmic microwave background anisotropies by the {\it Wilkinson Microwave Anisotropy Probe (WMAP)} satellite 
indicate that the Thomson optical depth to the last scattering surface is $\tau_e \sim 0.16$,
which in turn is suggestive of a high reionization redshift $z \sim 20$ (Kogut et al. 2003, Spergel et al. 2003).
Various scenarios have been put forth to account for these observations
which predict very different histories for the evolution of the ionization fraction
(e.g. Ciardi, Ferrara \& White 2003, Cen 2003, Wyithe \& Loeb 2003;
see Miralda-Escude 2003, Haiman 2003 for reviews).
GRBs may be the most promising tools to differentiate between these possibilities,
and some methods using near-infrared spectroscopy (Miralda-Escude 1998, Barkana \& Loeb 2003)
or photometry (Inoue, Yamazaki \& Nakamura 2003) have been proposed.
 
On an entirely different note,
the properties of the local environment in which GRBs explode is a crucial issue,
bearing on the still mysterious nature of the GRB progenitors.
At present, the density of the medium surrounding GRBs is quite uncertain,
even for the well-observed bursts at low redshifts.
Through self-consistent broadband modeling of a sample of bright GRB afterglows,
Panaitescu \& Kumar (2001, 2002) derive ambient densities in the range $n \sim 10^{-3}-30 {\rm cm^{-3}}$
on the scales of external shock deceleration (radii $r \sim 10^{16}-10^{18} {\rm cm}$).
In light of the numerous indications for massive star progenitors,
one may expect a denser, radially decreasing medium characteristic of stellar winds (Chevalier \& Li 1999),
but this is generally not supported by observations.
On the other hand, there is tentative (but far from conclusive) evidence
from soft X-ray absorption and optical/X-ray flux ratios that the GRB environment
may have higher densities characteristic of Galactic molecular clouds (MCs),
$n \sim 10^2-10^{3} {\rm cm^{-3}}$ or even larger
(Galama \& Wijers 2000, Reichart \& Price 2002, Piro 2002 and references therein).
The actual environment may have a complex density structure such as that of superbubbles (Scalo \& Wheeler 2001).
Clearly, one would like to have as many ways as possible to probe the nature of the circumburst medium.

We make the case here that dispersion delay in the radio emission of GRBs
may be capable of providing valuable insight into both of these issues.
\footnote{A similar study has recently been carried out by Ioka (2003).}
Dispersion is sensitive to the integrated column density of ionized material along the line of sight,
and is therefore unique and complementary to methods utilizing absorption by intervening neutral gas.
We show that dispersion delay may be effective in discriminating between different reionization histories,
particularly if GRBs emit strong, coherent low frequency emission as predicted by some recent models
(Sagiv \& Waxman 2002).
Interesting constraints may also be achievable for the ill-understood warm-hot IGM at low redshift
(Cen \& Ostriker 1999, Dav\'e et al. 1999, 2001).
Dispersion of GRB radio emission as a probe of the IGM was first discussed by Ginzburg (1973)
and later by Palmer (1993).
We adopt the following cosmological parameters that are consistent with {\it WMAP} results:
$\Omega_m=0.27$, $\Omega_\Lambda=0.73$, $h=0.71$ and $\Omega_b=0.044$ (Spergel et al. 2003).

This paper is organized as follows.
In \S \ref{sec:disigm}, we evaluate the degree of dispersion due to the ionized IGM
and the differences expected for various proposed reionization scenarios.
Dispersion due to different types of local ionized media and its evolution with redshift
is discussed in \S \ref{sec:disloc}.
\S \ref{sec:ffabs} addresses free-free absorption in these environments.
In \S \ref{sec:disgrb}, we discuss the observability of dispersion delay
in the known afterglow emission from GRBs as well as the predicted coherent emission.
We conclude in \S \ref{sec:conc} along with remarks on future prospects.
The formalism for calculating the radio afterglow emission is detailed in the Appendix.

\section{Dispersion in the Ionized Intergalactic Medium}
\label{sec:disigm}

An electromagnetic signal of frequency $\nu$ will propagate through an ionized medium
with a velocity $v=c(1-\nu_p^2/\nu^2)^{1/2}$, less than the speed of light in vacuum (e.g. Rybicki \& Lightman 1979).
Here $\nu_p=(n_e e^2/\pi m_e)^{1/2}$ is the plasma frequency and $n_e$ is the free electron density of the medium.
Therefore the arrival time of a low frequency signal will be delayed with respect to that at $\nu \gg \nu_p$
by a frequency-dependent amount which is proportional
to the integrated column density of free electrons along the propagation path,

\begin{eqnarray}
\Delta t (\nu) = {e^2 \over 2\pi m_e c} {1 \over \nu^2} \int n_e dl \ ,
\label{eq:dt}
\end{eqnarray}
where $dl$ is an element of distance along the path.
The dispersion measure (DM) $\int n_e dl$ is commonly expressed in units of ${\rm pc \ cm^{-3}}$.
Observations of dispersion in the radio emission of pulsars in our Galaxy
show that the DM of the Galactic disk is $\sim 100 {\rm pc \ cm^{-3}}$ for typical lines of sight,
with maximum values of $\sim 1000 {\rm pc \ cm^{-3}}$ for lines of sight near the Galactic center
(e.g. Cordes \& Lazio 2002).

The intergalactic medium (IGM) is known to be significantly ionized
out to at least $z \la 6$ (Gunn \& Peterson 1965, Djorgovski et al. 2001).
The radio emission from high redshift GRBs may then be affected
by dispersion in the ionized IGM (Ginzburg 1973, Palmer 1993).
Note that the discussion of this section may also apply to any other type of transient radio source.
For propagation through cosmological distances, the appropiate comoving distance element
is $dl = c \left|dt/dz\right| dz$,
where $\left|dt/dz\right|=(1+z)^{-1}H(z)^{-1}$ and $H(z)=(\Omega_m(1+z)^3+\Omega_\Lambda)^{1/2}$ for a flat cosmology.
\footnote{Strictly speaking, the propagation path will deviate from a null geodesic,
but this effect is negligibly small unless $v \ll c$.}
One must also account for the redshifting frequency and time dilation during propagation.
Thus, for a source at redshift $z$, the mean dispersion delay at observer frequency $\nu$ due to the IGM is
\begin{eqnarray}
\Delta t_{\rm IGM} (\nu, z) = {e^2 \over 2 \pi m_e c^2} {1 \over \nu^2} \int_{0}^{z} dz' c \left|{dt \over dz'}\right| {n_e(z') \over 1+z'} \ .
\label{eq:dtigm}
\end{eqnarray}
The mean free electron density is $n_e(z)=x_e(z) n_{e,0} (1+z)^3$, where $x_e(z)$ is the ionization fraction.
We limit the discussion here to the average effects of the ionized IGM assuming homogeneity;
the implications of a more realistic, inhomogeneous IGM structure is touched upon in \S \ref{sec:conc}.
Assuming that the IGM comprises $Y=0.24$ of helium by mass fraction and the rest hydrogen,
and that $x_e(z)$ of the hydrogen and helium are fully and singly ionized, respectively (e.g. Yoshida et al. 2003),
$n_{e,0} = 0.92 \times 10^{-5} \Omega_b h^2 {\rm cm^{-3}}$,
or $n_{e,0} = 2.1 \times 10^{-7} {\rm cm^{-3}}$ using the {\it WMAP} value of $\Omega_b h^2$.
(Helium may become fully ionized at $z \la 3$, but this is neglected
as it will only result in a small difference.)

\begin{figure}
\centering
\epsfig{file=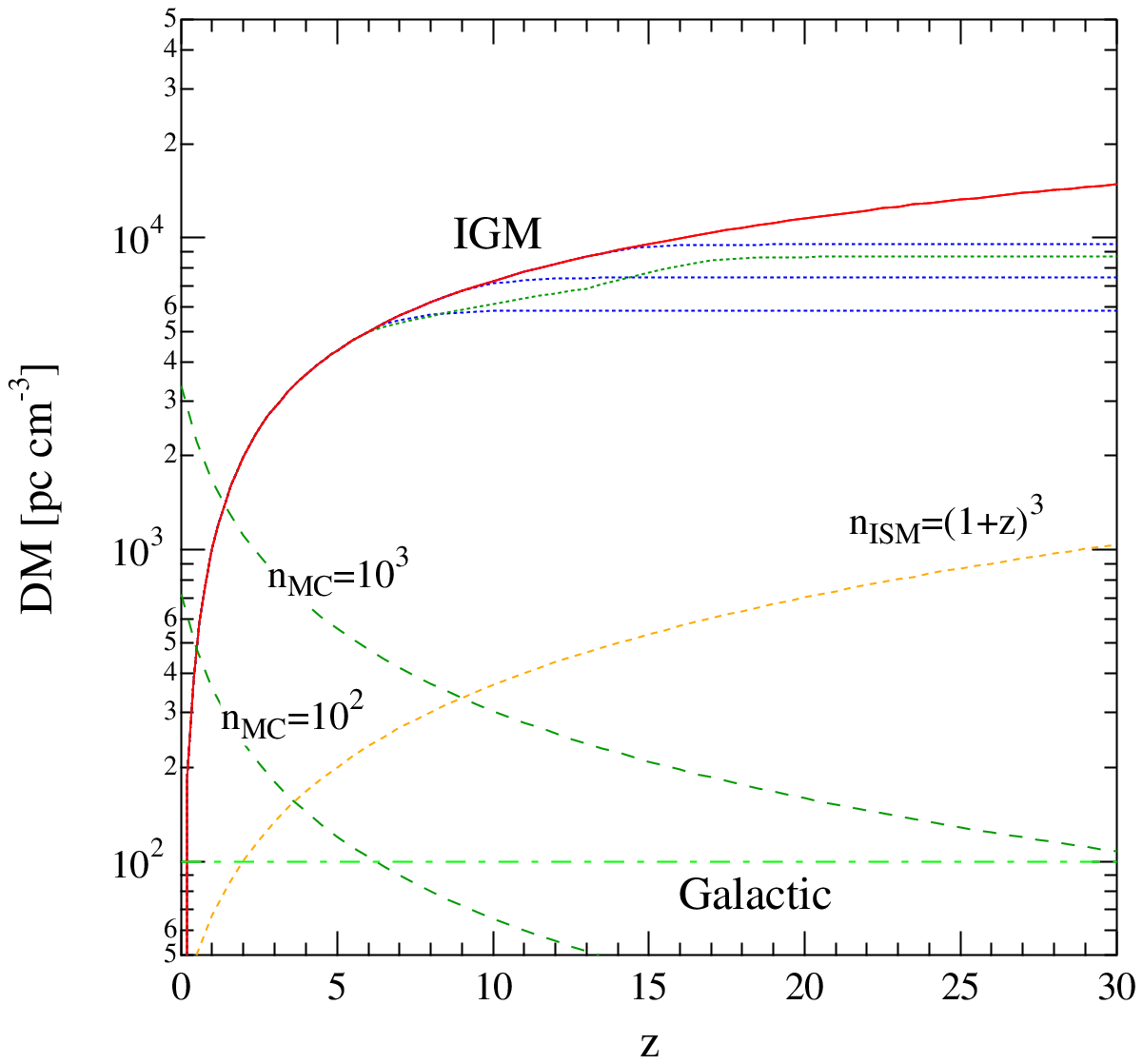,width=0.45\textwidth}
\epsfig{file=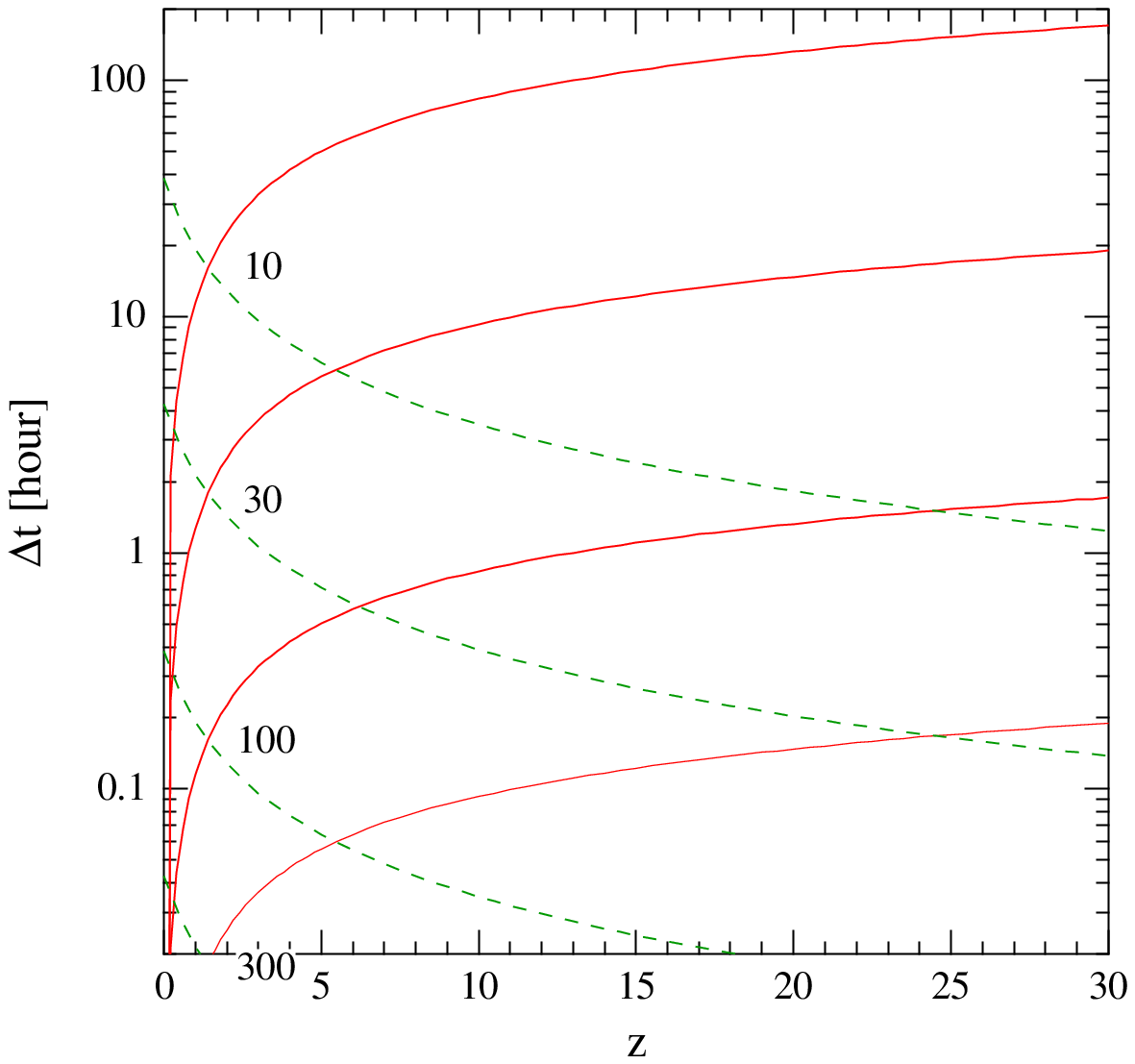,width=0.45\textwidth}
\caption{
a) Dispersion measures for different source redshifts due to the IGM (solid curve),
molecular cloud environments of
$n_{MC}=10^2 {\rm cm^{-3}}$ (lower dashed curve) and $n_{MC}=10^3 {\rm cm^{-3}}$ (upper dashed curve),
and the host galaxy ISM with $n_{ISM}=(1+z)^3 {\rm cm^{-3}}$ (dotted curve).
The curves ending horizontally at high $z$ correspond to different reionization histories
as discussed in the text and Fig.\ref{fig:dtreion}.
The typical Galactic dispersion measure (dot-dashed line) is also shown.
b) Dispersion delay times in hours
due to the IGM (solid curves) and MCs of $n_{MC}=10^3 {\rm cm^{-3}}$ (dotted curves),
for observer frequencies $\nu=$10, 30, 100 and 300 MHz, from top to bottom.
}
\label{fig:dmdt}
\end{figure}

The DM of the IGM out to different redshifts, defined appropriately as the integral in Eq.\ref{eq:dtigm},
is shown in Fig.\ref{fig:dmdt}a assuming for illustration that $x_e(z)=1$, i.e. the universe is always ionized.
We see that the IGM DM already exceeds typical Galactic values at redshifts as low as $z \sim 0.2$,
surpasses $\sim 1000 {\rm pc \ cm^{-3}}$ at $z \ga 1$, and can reach $\sim 10^4 {\rm pc \ cm^{-3}}$ at $z \ga 16$.
Fig.\ref{fig:dmdt}b shows the dispersion delay times at different observer frequencies.
At $\nu=$300, 100 and 30 MHz, the correspsonding delay times for sources at $z \sim 10$ are 
roughly 5 minutes, 1 hour and 10 hours, respectively.
Thus, dispersion delay may cause a potentially measurable imprint 
of propagation through the ionized IGM in the light curves of GRB radio emission,
as we discuss in \S \ref{sec:disgrb}.

We also see that the dependence of DM and $\Delta t$ with respect to $z$
becomes much milder for $z \ga 5$.
This results from the $\nu^{-2}$ dependence of dispersion and the associated $(1+z)^{-2}$ factor
in the integral of Eq.\ref{eq:dtigm}, which counteracts the $(1+z)^3$ increase in the density
and $(1+z)$ increase in time dilation:
at a given observer frequency, the higher the source redshift, the higher the rest-frame frequency,
and hence less dispersion.
This tendency is not so favorable for the purpose of probing the details of the reionization epoch,
for which the most interesting range of redshifts are $z \sim 5-20$.
Nevertheless, different possible reionization histories
can lead to observationally significant differences in $\Delta t$ and its $z$-dependence,
which may give us a crucial handle on discriminating between various scenarios for reionization.

To demonstrate this, we consider two different types of reionization scenarios
which have been discussed in light of the {\it WMAP} results.
First, we take a more conventional picture of reionization by early massive stars,
referring to the numerical simulation results of Ciardi, Ferrara \& White (2003) for concreteness 
(see also Miralda-Escud\'e 2003, Haiman 2003 and references therein).
Making relatively conservative assumptions about the properties of the ionizing stars
and their initial mass function (IMF), these models predict that
the universe is gradually ionized until an epoch of complete reionization $z_r$ is reached,
the value of which depends on the form of the IMF, production efficiency of ionizing photons,
and their escape fraction from the host galaxy.
The numerically calculated evolution of the ionization fraction is given in Fig.3 of Ciardi, Ferrara \& White (2003).
We schematically describe this as $x_e(z)=10^{-0.3(z-z_r)}$ for $z > z_r$ and $x_e(z)=1$ for $z \le z_r$,
and take the cases of $z_r$ =6, 9 and 13.5. The latter two cases correspond to their models S5 and L20,
which are consistent with the {\it WMAP} Thomson optical depth but not with the $z \ga 6$ quasar Gunn-Peterson troughs.
On the other hand, the additional case of $z_r$=6 considered here,
which may be realized with a lower photon production efficiency,
may be more concondant with the $z \ga 6$ quasar observations, but less with the {\it WMAP} results.

Second, we consider the scenario proposed by Cen (2003) (see also Wyithe \& Loeb 2003, Sokasian et al. 2003),
in which the universe is reionized twice:
first by very massive Population III stars, after which a period of partial recombination follows,
and then for the second time by Population II stars with a different IMF and photon production efficiency.
Fig.9 of Cen (2003) shows the ionization fraction in his model, which we schematically represent as
$x_e(z)=10^{-0.3(z-16)}$ for $z > z_1=16$, $x_e(z)=1$ for $13 \le z \le 16$,
$x_e(z)=10^{-0.3}$ for $6 < z < 13$ and $x_e(z)=1$ for $z \le z_2=6$.
Note that in the intermediate recombination period, the ionization fraction is still considerable,
but the amount of neutral hydrogen is enough to make the universe highly opaque to Ly$\alpha$ photons
(Cen 2003).

\begin{figure}
\centering
\epsfig{file=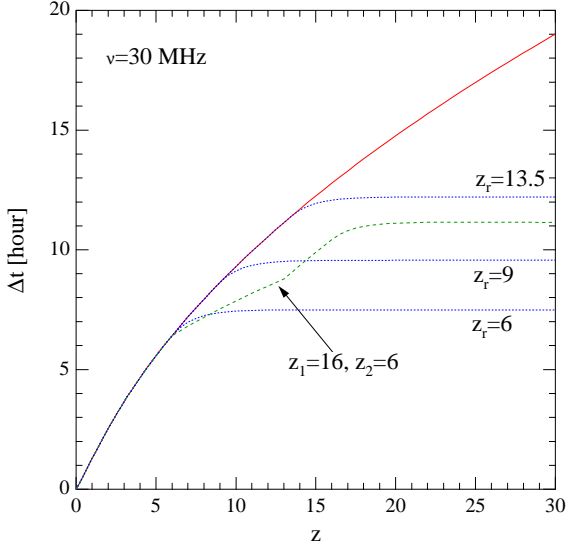,width=0.45\textwidth}
\caption{
The dispersion delay times in hours due to the IGM at $\nu=$30 MHz for scenarios with gradual reionization (dotted curves)
and two epoch reionization (short dashed curve). The case for a perpetually ionized IGM (solid curve)
is also shown.
}
\label{fig:dtreion}
\end{figure}

The dispersion delay times at $\nu$=30 MHz
for these different reionization histories are displayed in Fig.\ref{fig:dtreion} (note the linear vertical scale).
For the gradual reionization scenarios, $\Delta t$ keeps increasing with $z$
until the reionization redshift $z_r$ is reached, and remains constant thereafter at $z \ga z_r$.
For the two-epoch reionization scenario, the trend of $\Delta t$ with $z$
becomes shallower above the second reionization redshift
but does not completely level off as the recombination in the intermediate epoch is only partial.
The behavior steepens again from the second recombination redshift until the first reionization redshift,
above which $\Delta t$ becomes constant.

For $\nu$=30 MHz, $\Delta t$ at $z \sim 10-15$ is $\sim$ 10 hours,
and the the different reionization histories lead to differences in $\Delta t$ of several hours.
At $\nu$=100 MHz, these are $\sim 1$ hour and tens of minutes, respectively.
If the radio emission of GRBs at these frequencies are bright enough,
these delays ought to be measurable (see \S \ref{sec:coh}).
Measuring $\Delta t$ for a number of GRBs occurring at different redshifts,
and determining their redshifts independently,
e.g. through infrared Ly$\alpha$ breaks (Ciardi \& Loeb 2000, Gou et al. 2003),
X-ray lines (M\'esz\'aros \& Rees 2003) or 21 cm absorption features (Furlanetto \& Loeb 2003)
\footnote
{However, we find that radio afterglows from high-$z$ bursts
are typically too faint for detecting 21 cm absorption features due to the host galaxy; \S \ref{sec:after}.}
in their afterglow spectra,
we may be able to effectively discriminate between different reionization histories.
GRB radio emission may then serve as extremely valuable probes of the cosmic reionization epoch
which is otherwise difficult to investigate directly.

We mention that below $z \sim 3$, a progressively larger fraction of the IGM mass should be
incorporated into the warm-hot phase at temperatures $T \sim 10^5 - 10^7$ K,
being shock heated by infall onto incompletely virialized large scale structure such as filaments
(Cen \& Ostriker 1999, Dav\'e et al. 1999, 2001).
Whereas the average overdensity of the
cooler, photoionized IGM giving rise to the Ly$\alpha$ forest should be $\delta \sim 0$,
this warm-hot IGM (WHIGM) is expected to possess overdensities $\delta \sim 10-30$,
systematically above the mean baryon density of the universe (Dav\'e et al. 2001).
Although dispersion is insensitive to the medium temperature,
this difference in the average density of the WHIGM
may cause an increase in the dispersion above the simple prediction of Eq.\ref{eq:dtigm},
especially as the contribution to the DM from $z \la 3$ is very large (Fig.\ref{fig:dmdt}).
However, even though the mass fraction of the WHIGM can be
as high as 30-40 \% at $z \sim 0$ (5-20 \% at $z \sim 2$) (Dav\'e et al. 2001),
its volume fraction is expected to be much smaller,
less than 10 \% at $z \sim 0$ (5 \% at $z \sim 2$) (Cen \& Ostriker 1999).
Thus, the fraction of the line of sight to a GRB that is occupied by the higher density WHIGM
may not be so large.
Nevertheless, this contribution from the WHIGM may be important for some lines of sight,
possibly complicating studies of the reionization epoch with high-$z$ bursts.
On the other hand, given the difficulty of directly observing
this important but poorly understood component of the low redshift IGM
(compare e.g. Tripp, Savage \& Jenkins 2000, Mathur, Weinberg \& Chen 2003, Zappacosta et al. 2002),
interesting constraints may be obtainable through dispersion in GRBs at $z \la 3$
(provided that local dispersion is unimportant; \S \ref{sec:disloc}),
particularly if the degree of inhomogeneity can be quantified (see \S \ref{sec:conc}).
Further discussion of the effect of the WHIGM on dispersion delay is beyond the scope of this paper,
but may warrant a more detailed study.

\section{Dispersion in Ionized Local Environments}
\label{sec:disloc}

Besides the global IGM, there may be ionized regions locally enshrouding the GRB 
that can give rise to additional dispersion. Such local contributions may be described by
\begin{eqnarray}
\Delta t_{\rm loc} (\nu, z) = {e^2 \over 2\pi m_e c} {1 \over \nu^2} {N_e(z) \over 1+z} \ ,
\label{eq:dtloc}
\end{eqnarray}
where $N_e(z)$, the column density of free electrons local to the source, can in general evolve with $z$.
The DM here is $N_e(z)/(1+z)$.

Local ionized regions can come about in two ways:
i) photoionization of the GRB vicinity by UV photons from the GRB itself
(e.g. Perna \& Loeb 1998, Fruchter, Krolik \& Rhoads 2001, Draine \& Hao 2002, Perna \& Lazzati 2002), or
ii) photoionization (and/or mechanical heating) of the GRB surroundings by a population of massive stars
preexisting in the host galaxy (e.g. Ciardi et al. 2001, Barkana \& Loeb 2003).

In the second case, the HII regions may extend well out into the ambient IGM
for galaxies above the reionization redshift.
However, we can immediately see that
the dispersion from the local HII regions in the IGM should be generally miniscule.
Depending on the IMF and age of the massive star population,
the radii of these regions should be in the range $r_{HII} \sim 100-300$ kpc (Barkana \& Loeb 2003).
Since the IGM density is $n_{IGM} \simeq 2.1 \times 10^{-7} {\rm cm^{-3}} (1+z)^3$,
the DM is $n_{IGM} r_{HII}/(1+z) \simeq (2-6) \times 10^{-2} {\rm pc \ cm^{-3}} (1+z)^2$,
below the Galactic DM even at $z \sim 30$.
As one approaches the global reionization epoch, the HII regions become much larger
and begin to mutually overlap, but then the associated dispersion should be attributed to the global IGM.

Local dispersion can be important for much denser locations, as discussed below.
In such regions, preexisting massive stars play a lesser role
compared to GRB UV emission in photoionizing the environment,
since the hydrogen recombination time $t_{rec} \sim 10^5 {\rm yr} \ n^{-1}$
can be much shorter than the lifetime of the massive stars, $10^7-10^8$ yr.

\subsection{Molecular Clouds}
\label{sec:mc}

The current controversy regarding the environments of the observed GRBs 
and the circumstantial evidence for molecular clouds was described in \S \ref{sec:intro}.
As an exemplary case, we here consider the dispersion that may be caused
by environments similar to Galactic MCs, with typical size $r_{MC} \sim 10$ pc
and densities $n_{MC} \sim 10^2-10^{3} {\rm cm^{-3}}$ (e.g. Reichart \& Price 2002;
note that MC cores may have smaller sizes and higher densities).
Under the assumption that the GRB emits a flash of $10^{50} E_{UV,50}$ erg
in the UV (Draine \& Hao 2002; although the actual UV fluence may vary considerably from burst to burst;
see remark in \S \ref{sec:ffabs}),
these UV photons can ionize hydrogen with density $n$ out to a radius of roughly
\begin{eqnarray}
r_{ion} \sim 10^{20} {\rm cm} \ E_{UV,50}^{1/3} n^{-1/3} \ .
\label{eq:rion}
\end{eqnarray}

For $n_{MC} \sim 10^2-10^{3} {\rm cm^{-3}}$, $r_{ion} \sim 3-7$ pc.
If the MC properties do not evolve with $z$,
the corresponding DMs will be $\sim (720-3300) (1+z)^{-1} {\rm pc \ cm^{-3}}$.
Compared with the IGM DM and delay times in Fig.\ref{fig:dmdt}, we see that dispersion due to MCs
may dominate at low redshifts $z \la 1-2$, but becomes rapidly unimportant at higher $z$,
simply due to the $(1+z)^{-1}$ factor.
Thus, by measuring this effect for low-$z$ GRBs
one may gain valuable insight into the controversial issue of the GRB environment
(although free-free absorption may foil this prospect; see \S \ref{sec:ffabs}).
At the same time, such local contributions will not contaminate dispersion by the IGM in high-$z$ GRBs.

\subsection{Interstellar Medium of Host Galaxy}
\label{sec:hizism}

At high redshifts, the deficiency of metals and the associated reduction in cooling
may thwart the formation of MCs similar to the Galactic ones.
On the other hand, the average density of the interstellar medium (ISM) of host galaxies should increase with $z$,
possibly leading to another source of dispersion.
Lacking observational constraints, there is considerable uncertainty in how the ISM density evolves at high $z$.
The properties may be different for the metal-free Population III hosts 
and the more metal-enriched Population II hosts, the transition between the two
possibly taking place around $z \sim 10-15$ (Cen 2003, Wyithe \& Loeb 2003, Salvaterra \& Ferrara 2003).

Based on the disk galaxy models of Fall \& Efstathiou (1980) and Mo, Mao \& White (1998),
which are consistent with the observed properties of disks and damped Ly$\alpha$ absorbers at $z \la 3$,
Ciardi \& Loeb (2000) have discussed how the density of high $z$ disk galaxies may evolve.
In their model, the average ISM density and the disk scale height evolve respectively as
$n_{ISM} \propto M^{2/3} (1+z)^4$ and $h \propto M^{-1/3} (1+z)^{-2}$, where $M$ is the mass of the galaxy
(which is proportional to that of the dark matter halo).
Thus, for a fixed galaxy mass, the disk becomes very dense and thin at high $z$.
However, the standard picture of hierarchical structure formation implies that
the typical mass of galaxies should also evolve strongly.
For example, taking the minimum star-forming halo mass to be
$M_{\min} = 4.4 \times 10^9 M_\odot (1+z)^{-1.5} h^{-1}$ (Ciardi \& Loeb 2000),
the halo mass averaged over the Press-Schechter mass function evolves from 
$M \simeq 1.4 \times 10^{11} M_\odot$ at $z=0$ to $M \simeq 4.7 \times 10^8 M_\odot$ at $z=10$,
decreasing by more than two orders of magnitude.
The actual mass range relevant for GRB hosts should be that weighted by the amount of star formation,
the details of which is uncertain and model-dependent (e.g. Ciardi \& Loeb 2000, Barkana \& Loeb 2003).
In any case, if one is concerned with the {\em typical} GRB interstellar environment,
the density evolution in this model should be much milder than simply taking $(1+z)^4$ (c.f. Gou et al. 2003).

On the other hand, if the disk forms by collapsing by a constant factor in density
with respect to that of the halo, the ISM density should scale as $(1+z)^3$, as argued by Cen (2003).
The dependence on mass through the typical halo collapse redshift may be weak.
If we follow Cen (2003) in normalizing to the Galactic ISM density,
the density evolution in this case is $n_{ISM}=(1+z)^3 {\rm cm^{-3}}$.
This may possibly be consistent with numerical simulation results of primordial galaxy formation
which sometimes reveal disk-like structures with densities $n \sim 10^3-10^4 {\rm cm^{-3}}$ at $z \sim 20$
(Abel, Bryan \& Norman 2002, Bromm, Coppi \& Larson 2002, Yoshida et al. 2003).

In either situation, the DM due to the host galaxy ISM should be subdominant.
For the $(1+z)^3$ case, the ionization radius due to the GRB photons would be
$r_{ion} \simeq 33 {\rm pc} (1+z)^{-1}$,
so that the DM is $r_{ion} n_{ISM}(z)/(1+z) \simeq 33 {\rm pc \ cm^{-3}} (1+z)$, 
well below the IGM DM (Fig.\ref{fig:dmdt}).
Even if one makes the extreme (and unrealistic) assumption of $n_{ISM}= (1+z)^4 {\rm cm^{-3}}$,
the DM is $33 {\rm pc \ cm^{-3}} (1+z)^{5/3}$, which can reach $10^4 {\rm pc \ cm^{-3}}$ at $z \sim 30$
but is always overwhelmed by the IGM dispersion.

In addition, radiative and/or mechanical feedback, either by the GRB progenitor star or preexisting massive stars,
may be strong enough to keep the ambient densities as low as $n \sim 10^{-2}-1 {\rm cm^{-3}}$
(Scalo \& Wheeler 2001, Whalen, Abel \& Norman 2003), regardless of the average ISM density of the host galaxy,
further diminishing the DM.

Summarizing up to this point, molecular cloud environments can be the main contributor to dispersion
at low redshifts $z \la 1-2$, while the mean IGM should dominate at higher redshifts.
Other possible sources of dispersion are all expected to be minor.

\section{Free-Free Absorption}
\label{sec:ffabs}

Before turning to the observability of radio dispersion in GRBs,
we discuss the effect of free-free absorption,
which can strongly affect low frequency radio emission in dense, ionized environments,
as is the case for thermal emission from Galactic HII regions.
For an ionized medium consisting of pure hydrogen, density $n {\rm cm^{-3}}$,
temperature $10^4 T_4 K$ and depth $r_{pc}$ pc,
the optical depth to free-free absorption at observer frequency $\nu=10^8 \nu_8$ Hz is
\begin{eqnarray}
\tau_\nu = 5.5 \times 10^{-6} \nu_8^{-2} (1+z)^{-2} n^2 T_4^{-3/2} r_{pc} \bar{g}_{ff} \ ,
\label{eq:ffabs}
\end{eqnarray}
where the Gaunt factor appropriate for the radio regime is
$\bar{g}_{ff} = 0.28 \ln (T_4/\nu_8^{2}(1+z)^2) -7.2$ (Rybicki \& Lightman 1979, Shu 1991).

Free-free absorption by the IGM is completely negligible,
as $\tau_\nu \sim 10^{-8} \nu_8^{-2}$ even when assuming a constant temperature of $T_4=1$;
in reality the temperature should rise at higher $z$ (e.g. Carilli, Gnedin \& Owen 2002).
However, absorption can be serious for the dense local environments.
For the MCs discussed above with $n_{MC}=10^2-10^3 {\rm cm^{-3}}$ and $r_{ion} \sim 3-7$ pc,
$\tau_\nu \sim 0.7-30 \nu_8^{-2}$ at $z=1$.
For the host galaxy ISM with $n_{ISM}=(1+z)^3 {\rm cm^{-3}}$,
$r_{ion} \sim 33 {\rm pc} (1+z)^{-1}$, so that
$\tau_\nu = 1.4 \times 10^{-3} \nu_8^{-2} (1+z)^3 T_4^{-3/2} \bar{g}_{ff}$,
which is $\tau_\nu \sim 1.4 \nu_8^{-2}$ at $z=10$ and $\tau_\nu \sim 30 \nu_8^{-2}$ at $z=30$.
Under such circumstances, radio emission below 0.1-1 GHz can be strongly attenuated,
hampering the chances of observing dispersion effects at lower frequencies.
On the other hand, observing the spectral break due to free-free absorption
may be used to advantage as an alternative probe of dense, ionized environments.

One loophole is that the efficiency of flash photoionization of the GRB surroundings
may vary significantly from burst to burst,
as the fluence of UV ionizing photons emitted from the reverse shock
can depend strongly on the initial Lorentz factor, GRB duration and other parameters
(Kobayashi 2000, Waxman \& Draine 2000, Draine \& Hao 2002, Perna \& Lazzati 2002).
In the case of inefficient photoionization by the GRB, neither free-free absorption nor dispersion
by the local environment will affect the radio emission,
unless the prior activity of nearby massive stars have been vigorous.
Dispersion by the IGM should be the only effect of observational consequence.

\section{Dispersion Delay in GRB Radio Emission}
\label{sec:disgrb}

We now discuss the observability of dispersion delay effects in the radio emission from GRBs.

\subsection{Radio Afterglows}
\label{sec:after}

We first consider the known radio afterglow emission,
which can be well-described as incoherent synchrotron emission from electrons
accelerated by a relativistic blastwave decelerating in the ambient medium
(Paczy\'nski \& Rhoads 1993, Katz 1994, M\'esz\'aros \& Rees 1997, Sari, Piran \& Narayan 1998;
see Piran 1999, van Paradijs et al. 2000, M\'esz\'aros 2002 for reviews).
The calculation of the spectra and light curves follows standard discussions and is detailed in the Appendix.
Particular attention has been made to account correctly for synchrotron self-absorption
at high ambient densities.
Although radio emission is expected from both the long-lived forward shock
and the shorter-lived reverse shock (Sari \& Piran 1999, Kobayashi \& Sari 2000),
we neglect the latter component as we find it to be important only above a few GHz.
At the lower frequencies of observational interest for dispersion effects,
the forward shock emission is always dominant.

The afterglow flux at radio frequencies depends strongly on the density of the ambient medium,
especially in the self-absorbed part of the spectrum.
Much of the discussion of \S \ref{sec:disloc} regarding the density of the GRB environment
and its high $z$ evolution is also relevant here.
However, we note that the ambient density on the scales of external shock deceleration
(radii $10^{16}-10^{18}$ cm) may or may not be different from that contributing to the dispersion
(and free-free absorption).
For example, the external shock may happen to decelerate in a location of low density
while the afterglow emission passes through a much denser region in the foreground toward the observer.
This may occur in a complicated geometry like that of superbubbles (Scalo \& Wheeler 2001),
or even for Population III conditions if feedback from the progenitor star
creates a low density neighborhood within a denser ISM (Whalen et al. 2003).
We will refer to the medium relevant to shock deceleration as the circumburst medium (CBM),
and denote its density as $n_{CBM}$.

\begin{figure}
\centering
\epsfig{file=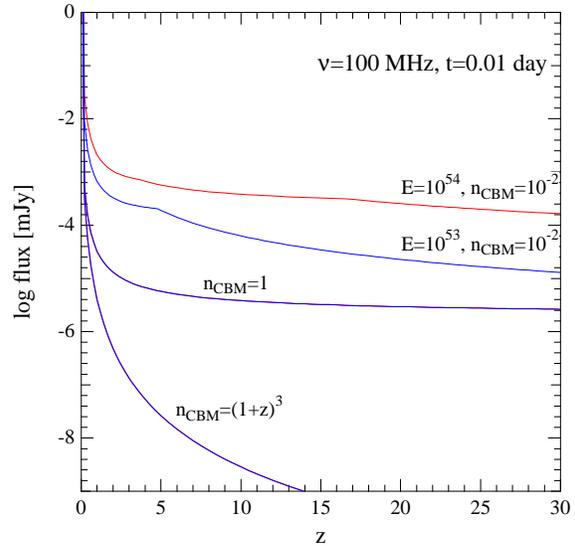,width=0.45\textwidth}
\caption{
The dependence on $z$ of the afterglow flux at $\nu=100$ MHz and $t=0.01$ day
(not including dispersion or free-free absorption),
for different assumptions on $E$ and $n_{CBM}$ as labeled.
The two lower cases are independent of $E$.
}
\label{fig:rfz}
\end{figure}

Considering three different cases,
$n_{CBM}=1$, $10^{-2}$ and $(1+z)^3$, all in units of ${\rm cm^{-3}}$,
Fig.\ref{fig:rfz} plots the $\nu=100$ MHz flux at a fixed time $t=0.01$ day ($\sim$ 15 min) after the burst,
not including the effects of dispersion or free-free absorption.
At this range of $\nu$ and $t$, typically the spectra are rising with $\nu$
and the light curves are rising with $t$.
Thus, two effects compete in the dependence of the flux on $z$:
time dilation makes one sample the light curve at an earlier time with less flux,
whereas K-correction makes one observe the spectrum at higher frequencies with more flux.
This can result in a relatively flat behavior if $n_{CBM}$ does not evolve,
sometimes with a kink when the self-absorption frequency is crossed.
In constrast, the case of $n_{CBM} \propto (1+z)^3$ causes such a strong increase in
the self-absorption that the flux is suppressed rapidly to unobservable levels above $z$ of a few.
The isotropic equivalent (i.e. beaming-corrected)
initial kinetic energy of the blastwave is fiducially $E=10^{53}$ erg
from the most recent observational determinations
(Panaitescu \& Kumar 2001, 2002, Frail et al. 2001).
We also show the case of an exceptionally bright GRB with $E=10^{54}$ erg,
but the difference is apparent only for $n_{CBM}=10^{-2} {\rm cm^{-3}}$.
For either $n_{CBM}=1 {\rm cm^{-3}}$ or $(1+z)^3  {\rm cm^{-3}}$,
the flux is always in the self-absorbed regime,
in which case it is actually independent of $E$.
Even in the most optimistic (and rare) situation
of $E=10^{54}$ erg and $n_{CBM}=10^{-2} {\rm cm^{-3}}$,
the flux is at $\mu$Jy levels, requiring powerful telescopes for detection.

\begin{figure}
\centering
\epsfig{file=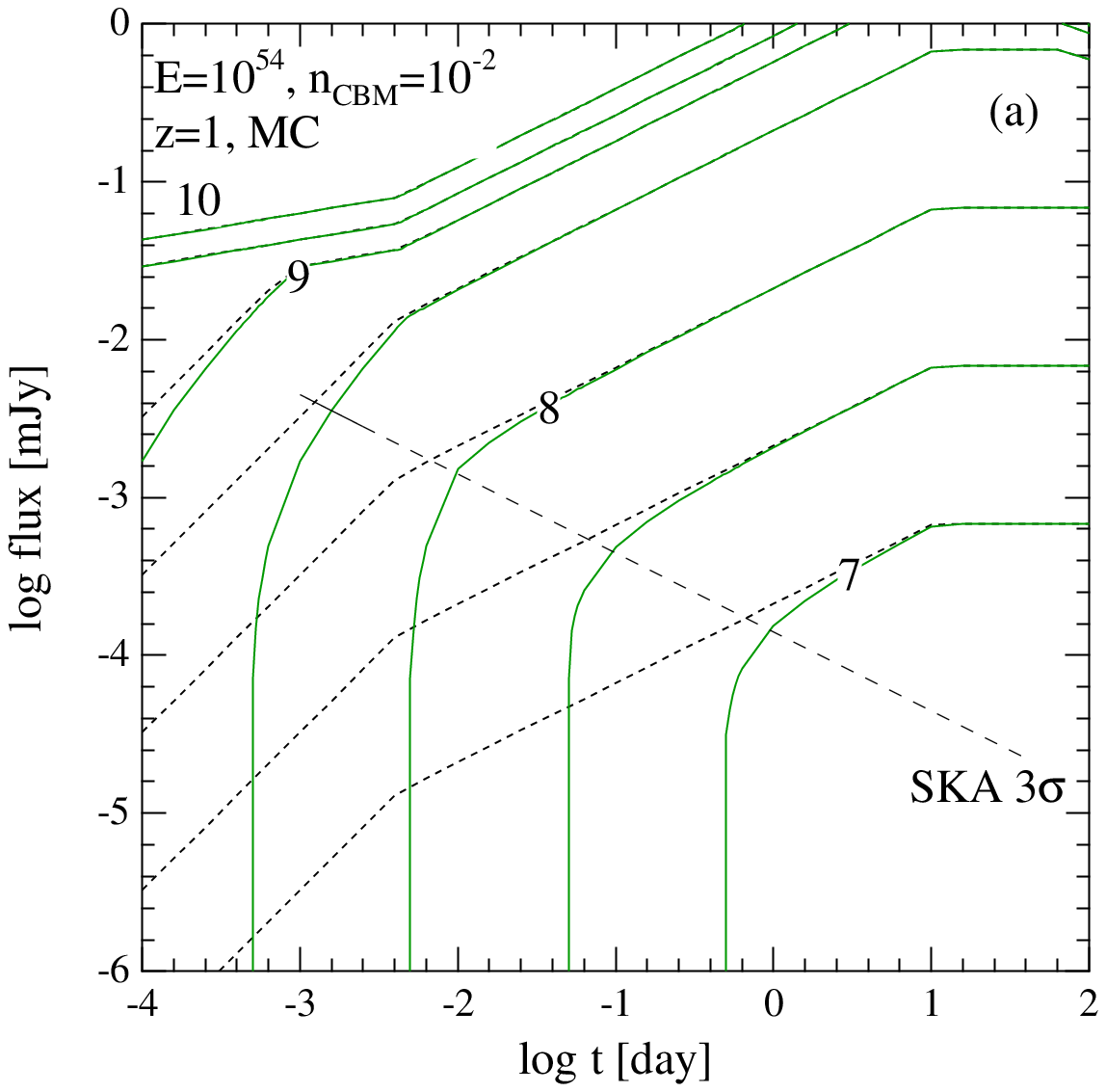,width=0.45\textwidth}
\epsfig{file=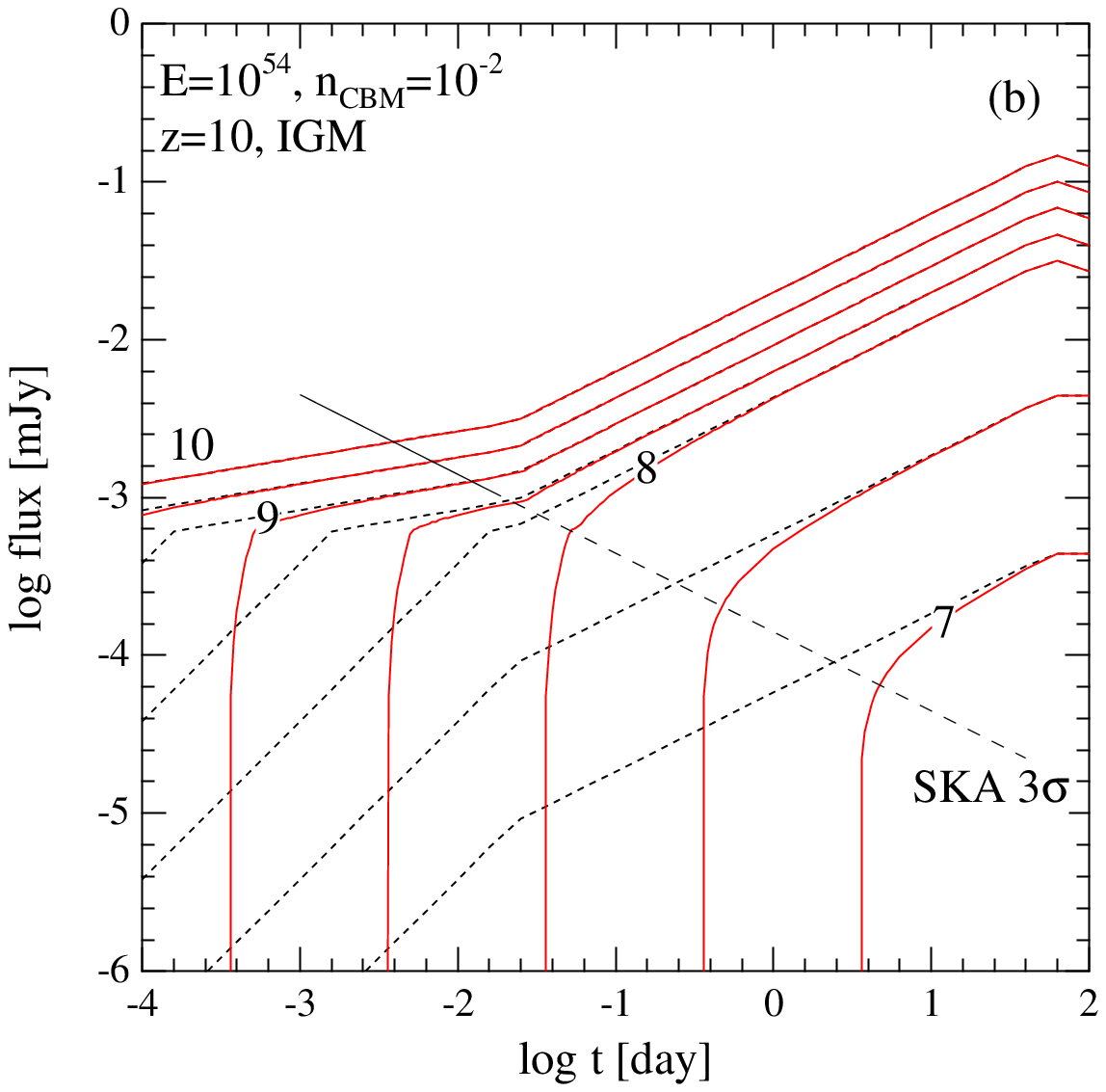,width=0.45\textwidth}
\caption{
Light curves of radio afterglows for $E=10^{54}$ erg and $n_{CBM}=10^{-2} {\rm cm^{-3}}$,
at frequencies $\nu =$ 10, 30, 100, 300 MHz, 1, 3 and 10 GHz, from bottom to top,
with (solid curves) and without (dashed curves) dispersion.
a) For the case of intervening molecular cloud of $n_{MC}=10^3 {\rm cm^{-3}}$ at $z=1$,
neglecting free-free absorption.
b) For the case of ionized IGM at $z=10$.
Overlayed is the 3 $\sigma$ sensitivity of {\it SKA} assuming an integration time 30\% of $t$.
The dashed portion indicates the range where dispersion effects are actually unobservable
due to the {\it SKA} low frequency limit of 150 MHz.
}
\label{fig:rlc}
\end{figure}

Dispersion simply delays the light curve at observer time $t$ and frequency $\nu$
from $f(\nu,t)$ to $f(\nu, t-\Delta t(\nu))$.
Concentrating on the most optimistic parameters above, we display in Fig.\ref{fig:rlc}
the radio light curves at different frequencies, with and without dispersion for two cases.
One is for a $z=1$ burst with dispersion by an intervening MC of $n_{MC}=10^3 {\rm cm^{-3}}$
(which the GRB must reside outside of, but near enough so as to be able to sufficiently photoionize).
In this case, free-free absorption by the MC should be strong in reality,
but we do not account for this here in order to isolate the consequences of dispersion.
The other case is for a $z=10$ burst with dispersion by the ionized IGM.
Overlayed is the 3-$\sigma$ sensitivity of the {\it Square Kilometer Array (SKA)}
\footnote{http://www.skatelescope.org},
the most sensitive radio telescope currently being conceived,
assuming that the integration time is 30 \% of the observed time.

We see that the effect is to strongly suppress the early, power-law rise of the light curves
until the delay time $\Delta t$, after which the flux rises more sharply than expected without dispersion.
Observationally distinguishing this difference is in principle possible,
but is likely to be a daunting task.
The lowest frequency band of the current {\it SKA} design is at $\nu=150$ MHz,
at which $\Delta t \sim$ 5 minutes for the MC case and $\sim$ 20 minutes for the IGM case.
Satellite missions such as {\it SWIFT} \footnote{http://swift.gsfc.nasa.gov} should be able
to localize GRB positions to arcminute accuracy 
and distribute them within about a minute of the burst.
The telescope response time of {\it SKA} can be short enough to point
and begin observing the GRB position within another minute or so.
By setting sufficient upper limits at the earliest times
and characterizing the delayed, sharp rise of the light curve,
one may be able to determine the delay time.
The case may be strengthened if this can be done at one or more neighboring frequency bands
so far as $\Delta t$ is more than a few minutes.
However, establishing the absence of a signal near the telescope sensitivity limit
is undoubtedly very difficult.
The early radio light curve should also be significantly modulated by scintillation
due to the Galactic ISM (Goodman 1997, Waxman, Frail \& Kulkarni 1998),
and may be contaminated by emission from the host galaxy as well.
Most seriously, confusion with other sources in the field of view
(or even within the angular resolution element of {\it SKA}; Garrett 2002)
is likely to be severe at $\mu$Jy.
Hope for distinguishing between different cosmic reionization histories in this way is remote.
We conclude that detecting the effect of dispersion in the radio afterglow emission
should be extremely challenging but may be marginally possible,
only for rare, bright GRBs occurring in low density environments.

\subsection{Coherent Radio Emission}
\label{sec:coh}

The incoherent synchrotron emission of radio afterglows is strongly self-absorbed and faint at low frequencies,
limiting the detectability of the dispersion signature. 
However the observational prospects can be much better if GRBs emit strong, coherent radio emission,
as predicted by some recent models.
Very different types of emission from a variety of physical processes have been suggested
(e.g. Benz \& Paesold 1998, Usov \& Katz 2000, Hansen \& Lyutikov 2001, Sagiv \& Waxman 2002).
The current observational limits on such emission associated with the prompt GRB is quite weak,
in the 10-100 kJy range, due to the limited sky coverage of current facilities
and difficulties in filtering out terrestrial interference
(e.g. Dessene et al. 1996, Katz et al. 2003 and references therein).

Particularly interesting for our purposes is the work of Sagiv \& Waxman (2002).
\footnote{Other models predict emission which is either too faint (Benz \& Paesold 1998, Hansen \& Lyutikov 2001)
or at too low frequency (Usov \& Katz 2000).}
Within the framework of the standard external shock picture of afterglows,
they demonstrate that under certain conditions, 
strong, maser-type coherent emission can emerge
on top of the incoherent, self-absorbed synchrotron emission
at the onset of shock deceleration, from either the forward or reverse shock.
For plausible energy distributions of the injected electrons,
this effect can be substantial provided that
the ratio of the postshock magnetic field energy to electron energy is $\la 10^{-5}$,
and the CBM is sufficiently dense, $n_{CBM} \sim 10^4 {\rm cm^{-3}}$.
The maser emission is characteristically narrow-band around the generalized Razin-Tsytovich frequency,
which in the case of the forward shock in a uniform CBM of $n_{CBM} = 10^4 n_{CBM,4} {\rm cm^{-3}}$ is
\begin{eqnarray}
\nu_R* \simeq 220 \ l_4^{1/4} \epsilon_{e,-1}^{-1/4} \epsilon_{B,-6}^{-1/4} E_{53}^{1/8} n_{CBM,4}^{3/8} T_{GRB,1}^{-3/8} {\rm MHz} \ ,
\label{eq:nur}
\end{eqnarray}
where $T=10 T_{GRB,1}$s is the burst duration, $l=4 l_4$ is a factor which depends on the electron distribution,
and other symbols are as defined in the Appendix (see Sagiv \& Waxman 2002 for details).
Although detailed predictions for the spectrum and light curve of this component are not yet available,
the emission should be generally correlated with the prompt GRB with durations of $\sim$ 1 minute,
and can be as bright as $\sim$ 1 Jy for low redshift bursts (E. Waxman, private communication).

The characteristic frequency of a few 100 MHz in the rest-frame
is just where the effects of dispersion delay
are observationally most interesting for probing the reionization epoch around $z \sim 10$,
which will be observed at few 10s of MHz (\S \ref{sec:disigm}; Fig.\ref{fig:dtreion}).
The bright peak flux and fast decaying light curve should allow the dispersion signal
to be readily distinguished, in the usual manner employed for pulsars (e.g. Cordes \& McLaughlin 2003).
Note that this emission is stronger for a denser CBM, in constrast
to the opposite trend for the self-absorbed incoherent emission.
The CBM may naturally be as dense as $10^4 {\rm cm^{-3}}$
for the ISM of high redshift galaxies or inside molecular cloud cores (\S \ref{sec:disloc});
a stellar-wind CBM is not necessary.
Free-free absorption by GRB photoionized material may not be a worry here:
since both the maser emission and UV flash peak at the same time,
and the forward shock is located in front of the reverse shock,
the propagation front of the maser emission should lead the photoionization front
(although the details may depend on the differences in the actual light curves).
The requirement of the low magnetic field may seem more stringent but is not precluded,
since the mechanism for amplifying the magnetic field above the compressed ISM value
behind the forward shock is still uncertain.
(Values obtained from afterglow modeling are in the range $\epsilon_B \sim 10^{-5}-10^{-1}$;
Panaitescu \& Kumar 2002).
In fact, very low magnetic fields may be inevitable for the ISM of high redshift
(especially Population III) galaxies (Abel et al. 2002, Bromm et al. 2002).
All these factors point to this maser emission being a valuable probe of the reionization epoch.

One should first observationally confirm the existence of such an emission component from low-$z$ GRBs
before it can be utilized as a cosmological probe with high-$z$ bursts.
The best-suited instrument for both objectives may be the {\it Low Frequency Array (LOFAR)},
with its high sensitivity at low frequencies, wide field of view, rapid response
and effective rejection capability of terrestrial signals
(R\"ottgering et al. 2002).
According to the current design, the sensitivity of the whole array
at 30 MHz is 68 mJy for an integration time of 1s and a bandwidth of 4 MHz
\footnote{http://www.lofar.org}.
Assuming that the maser emission has a redshifted duration of 10 min
(corresponding to a rest-frame duration of $\sim$ 1 min at $z \sim 10$)
in the frequency range $\Delta \nu \sim \nu$,
and that matched filters accounting for the dispersion delay (Cordes \& McLaughlin 2003)
have a detection efficiency of 30\%,
{\it LOFAR} should be able to detect this emission at 30 MHz down to $\sim$ 3 mJy,
provided that terrestrial and ionospheric interference signals at these fluxes can be discriminated.
If the mean flux at $z \sim 1$ is 1 Jy, this should be detectable out to $z \sim 10$ (delayed by 10 hours!).
Higher sensitivities and limiting redshifts may be achievable at higher frequencies
as long as that frequency falls in the redshifted emission band,
and the dispersion delay should be measurable 
as long as the delay time is longer than the redshifted duration of emission.
By collecting such information for a sufficiently large sample of GRBs at different redshifts
and determining their redshifts by other methods (Ciardi \& Loeb 2000, Gou et al. 2003, M\'esz\'aros \& Rees 2003),
one may be able to pin down the actual cosmic reionization history (Fig.\ref{fig:dtreion}),
which would be a great step forward in elucidating the most poorly understood era in the early universe.

\section{Conclusions and Prospects}
\label{sec:conc}

We summarize the main points of this work.
Dispersion should be dominated by the ionized IGM for sources at redshifts $z \ga 2$,
providing a valuable probe of the cosmic reionization history at high redshifts,
and possibly also the low redshift warm-hot IGM.
At $z \la 2$, dispersion by local material ionized by the GRB may be important in molecular cloud environments,
although free-free absorption can strongly attenuate the emission at the relevant frequencies.
For the standard radio afterglow emission, dispersion suppresses the flux in the early-time light curve,
the detection of which should be very difficult,
but perhaps marginally possible for rare, bright afterglows with sensitive telescopes such as {\it SKA}.
Dispersion delay should be much more readily detectable
if GRB emit prompt, coherent radio emission by the synchrotron maser mechanism
as discussed by Sagiv \& Waxman (2002).
{\it LOFAR} may be capable of detecting such emission out to $z \sim 10$ or higher,
potentially allowing for effective discrimination of different reionization scenarios.

Dispersion is a unique probe of the IGM in that it is sensitive to ionized matter,
as opposed to the neutral matter probed by most other methods utilizing absorption processes.
The main disadvantage is that it can only measure the total integrated column density of ionized gas to the source,
and cannot separate the contributions from different points along the line of sight to the observer.
However, we have seen that the high-$z$ dispersion signature reflects the global IGM
and is unlikely be contaminated by other local contributions, a crucial asset
for probing the cosmic reionization history (compare e.g. Barkana \& Loeb 2003).

In this work, we have only addressed the average effects of a homogeneous IGM.
In reality, the ionized IGM should have significant structure, due to
1) density fluctuations resulting from structure formation, i.e. gravitational evolution,
which becomes increasingly prevalent at lower redshifts (Cen \& Ostriker 1999, Dav\'e et al. 1999, 2001), and
2) variations in the ionization fraction from the inhomogeneous distribution of the ionizing sources,
which should have a highly complicated topology before the global reionization redshift (e.g. Ciardi et al. 2000).
One obvious consequence of the inhomogeneity is to introduce a scatter in the dispersion delay
around the mean expectation Eq.\ref{eq:dtigm} for sources along different lines of sight
(or a systematic increase for some lines of sight with large covering fraction by the warm-hot IGM;
\S \ref{sec:disigm}).
Of additional interest may be scattering effects:
radiation propagating in an ionized medium with varying density will have
its path bent due to the varying refractive index (Rybicki \& Lightman 1979).
Fluctuations in the free electron density along the propagation path
can lead to temporal broadening of a transient pulse,
as well as angular broadening of the image (e.g. Rickett 1990, Cordes \& Rickett 1998).
The magnitude of these effects depends on the spatial distribution of the ionized regions
and the power spectrum of the density fluctuations within them.
Estimating their importance for propagation through the inhomogeneous IGM is nontrivial (e.g. Goodman 1997),
and may require comparison with numerical simulation results.
If these effects turn out to be observationally interesting,
further valuable information concerning the density and ionization structure of the IGM
may be obtainable through the radio emission of GRBs.

\section*{Acknowledgments}
I acknowledge very informative discussions with B. Ciardi and T. En\ss lin.
Thanks also go to G. Bjornsson, R. Blandford, E. Churazov, A. Fruchter, R. Sunyaev and E. Waxman for helpful comments.
This work was partially supported by the European Research and Training Network
``Gamma-Ray Bursts: An Enigma and A Tool''.

\appendix

\section{Radio Afterglow Spectra and Light Curves}

We follow standard prescriptions for the synchrotron emission from electrons
accelerated by a relativistic shock decelerating in the ambient medium
(Sari, Piran \& Narayan 1998, Wijers \& Galama 1999, Panaitescu \& Kumar 2000),
with particular attention to self-absorption (see also Granot \& Sari 2002).
Aside from the ambient density $n {\rm cm^{-3}}$, here taken to be spatially uniform,
the principal parameters are fiducially taken to be consistent with
those obtained from broadband modeling (Panaitescu \& Kumar 2001, 2002):
the isotropic equivalent, initial kinetic energy of the blastwave is $10^{53} E_{53}$ erg,
the half angle of jet collimation is $0.1 \theta_{-1}$ rad
(Sari, Piran \& Halpern 1999; although this is not essential for this work),
the fractions of postshock energy imparted to relativistic electrons and magnetic fields
are $0.1 \epsilon_{e,-1}$ and $0.01 \epsilon_{B,-2}$ respectively,
and the spectral index of the electron distribution is $p=2.5$.
We assume an adiabatic shock, for which the relation between energy $E$,
bulk Lorentz factor $\Gamma$ and shock radius $r$ is $E=4\pi\Gamma^2 r^3 n m_p c^2/3$,
and adopt the relation for the observer time $t = r(1+z)/4 \Gamma^2 c$.

The electron distribution is a broken power-law,
with an injection break at Lorentz factor $\gamma_m = \epsilon_e \Gamma m_p(p-2)/m_e(p-1)$
and a cooling break at Lorentz factor $\gamma_c = 6 \pi m_e c/\Gamma \sigma_T B^2 t(1+Y)$,
where $B=(32 \pi \epsilon_B m_p n)^{1/2} \Gamma c$.
The Compton parameter $Y$ is $Y=(\sqrt{4\epsilon_e/\epsilon_B+1}-1)/2$
if the electrons are fast cooling ($\gamma_c < \gamma_m$),
while it decreases slowly with $t$
if the electrons are slow cooling ($\gamma_m < \gamma_c$) (Sari \& Esin 2001);
we approximate by taking the fast cooling value at all times.
The power-law index $q$ between these breaks is $q=2$ for fast cooling
and $q=p$ for slow cooling.
The observed characteristic synchrotron frequency associated with an electron of Lorentz factor $\gamma$
is $\nu=\Gamma eB \gamma^2/ \pi m_e c (1+z)$,
and the emitted synchrotron spectrum has breaks at $\nu_m$ and $\nu_c$ corresponding to $\gamma_m$ and $\gamma_c$.
The flux peaks at $\nu_p=\min(\nu_m, \nu_c)$ and the break frequency above it is $\nu_b=\max(\nu_m, \nu_c)$.
The peak flux at $\nu_p$ for a source at luminosity distance $D$ is
$f_{\nu,p}=(e^3/3^{1/2} m_e c^2) \Gamma B r^3 (1+z)/D^2$.

The evaluation of the synchrotron self-absorption frequency
is based on Granot, Piran \& Sari (1999) and Panaitescu \& Kumar (2000).
Using the self-absorption coefficient averaged over an isotropic distribution of pitch angles (Granot et al. 1999),
the self absorption optical depth at $\nu_p$ is
\begin{eqnarray}
\tau_p = \psi f(p) {e n r \over 3 B \gamma_p^5} \ ,
\label{eq:taup}
\end{eqnarray}
where $\gamma_p=\min(\gamma_m, \gamma_c)$, $f(p)=(p+2)(p-1)/(3p+2)$
and $\psi=2^{8/3} \pi^{5/2}/5 \Gamma(5/6)$.
This gives, for fast cooling,
\begin{eqnarray}
\tau_p = \tau_c = 0.79 \times 10^{-11} f(p) (1+Y)^5 \epsilon_{B,-2}^{9/2} E_{53}^2 n^{7/2} \ ,
\label{eq:tauc}
\end{eqnarray}
while for slow cooling,
\begin{eqnarray}
\tau_p &=& \tau_m = 3.5 \times 10^{-7} f(p) \left({p-1 \over p-2}\right)^5 \\ \nonumber
       &\times& \epsilon_{e,-1}^{-5} \epsilon_{B,-2}^{-1/2} E_{53}^{-1/2} n \left({t_d \over 1+z}\right)^{5/2} \ ,
\label{eq:taum}
\end{eqnarray}
where $t_d$ is the observer time in days.
The optical depth at an arbitrary frequency is then
\begin{eqnarray}
\tau_\nu &=& \tau_p
\left\{
\begin{array}{ll}
\displaystyle{\left(\nu \over \nu_p\right)^{-5/3}} & \nu < \nu_p \\
\displaystyle{\left(\nu \over \nu_p\right)^{-(q+4)/2}} & \nu_p < \nu < \nu_b \\ 
\displaystyle{\left(\nu \over \nu_b\right)^{-(p+5)/2} \left(\nu_b \over \nu_p\right)^{-(q+4)/2}} & \nu_b < \nu \\
\end{array}
\right. \
\label{eq:taunu}
\end{eqnarray}
The frequency at which $\tau_\nu=1$ gives the synchrotron self absorption frequency $\nu_a$,
\begin{eqnarray}
\nu_a &=& \nu_p
\left\{
\begin{array}{ll}
\displaystyle{\tau_p^{3/5}} & \tau_p < 1 \\
\displaystyle{\tau_p^{2/(q+4)}} & \tau_b < 1 < \tau_p \\ 
\displaystyle{\tau_p^{2/(p+5)} \left(\nu_b \over \nu_p\right)^{1-(q+4)/(p+5)}} & \tau_b > 1\\
\end{array}
\right. \ ,
\label{eq:nua}
\end{eqnarray}
where $\tau_b=\tau_p (\nu_b/\nu_p)^{-(q+4)/2}$ is the optical depth at $\nu_b$.
The case of $\tau_b > 1$ can become important for ambient densities $n \ga 10^4 {\rm cm^{-3}}$.

The spectra $f_\nu$ are broken power-laws with different slopes between the different characteristic frequencies.
Three cases can be distinguished depending on the location of $\nu_a$ with respect to $\nu_p$ and $\nu_b$:
for $\nu_a < \nu_p$,
\begin{eqnarray}
f_\nu &=& f_{\nu,p}
\left\{
\begin{array}{ll}
\displaystyle{\left(\nu \over \nu_a\right)^2 \left(\nu_a \over \nu_p\right)^{1/3}} & \nu < \nu_a \\
\displaystyle{\left(\nu \over \nu_p\right)^{1/3}} & \nu_a < \nu < \nu_p \\
\displaystyle{\left(\nu \over \nu_p\right)^{-(q-1)/2}} & \nu_p < \nu < \nu_b\\
\displaystyle{\left(\nu \over \nu_b\right)^{-p/2} \left(\nu_b \over \nu_p\right)^{-(q-1)/2}} & \nu_b < \nu\\
\end{array}
\right. \ ,
\label{eq:fnu1}
\end{eqnarray}
for $\nu_p< \nu_a < \nu_b$,
\begin{eqnarray}
f_\nu &=& f_{\nu,p}
\left\{
\begin{array}{ll}
\displaystyle{\left(\nu \over \nu_p\right)^2 \left(\nu_p \over \nu_a\right)^{5/2}} \left(\nu_a \over \nu_p\right)^{-(q-1)/2} & \nu < \nu_p \\
\displaystyle{\left(\nu \over \nu_a\right)^{5/2} \left(\nu_a \over \nu_p\right)^{-(q-1)/2}} & \nu_p < \nu < \nu_a \\
\displaystyle{\left(\nu \over \nu_p\right)^{-(q-1)/2}} & \nu_a < \nu < \nu_b\\
\displaystyle{\left(\nu \over \nu_b\right)^{-p/2} \left(\nu_b \over \nu_p\right)^{-(q-1)/2}} & \nu_b < \nu\\
\end{array}
\right. \ ,
\label{eq:fnu2}
\end{eqnarray}
and for $\nu_b < \nu_a$,
\begin{eqnarray}
f_\nu &=& f_{\nu,p}
\left\{
\begin{array}{ll}
\displaystyle{\left(\nu \over \nu_p\right)^2 \left(\nu_p \over \nu_a\right)^{5/2} \left(\nu_a \over \nu_b\right)^{-p/2}}\\
 \ \ \ \ \ \times \displaystyle{\left(\nu_b \over \nu_p\right)^{-(q-1)/2}} & \nu < \nu_p \\
\displaystyle{\left(\nu \over \nu_a\right)^{5/2} \left(\nu_a \over \nu_b\right)^{-p/2} \left(\nu_b \over \nu_p\right)^{-(q-1)/2}} & \nu_p < \nu < \nu_a\\
\displaystyle{\left(\nu \over \nu_b\right)^{-p/2} \left(\nu_b \over \nu_p\right)^{-(q-1)/2}} & \nu_a < \nu\\
\end{array}
\right. \ .
\label{eq:fnu3}
\end{eqnarray}
For fast cooling, an additional break may exist at $\nu < \nu_a$
due to the inhomogeneous electron distribution behind the shock front (Granot, Piran \& Sari 2000),
but this is neglected as it results in only a small change of the slope.

The light curves at a fixed frequency $\nu$ can show several breaks of different origin:
the passage of some characteristic frequency ($\nu_a$, $\nu_p$ or $\nu_b$) through $\nu$,
the transition from fast cooling to slow cooling,
and the jet transition due to edge visibility and sideways expansion.
A further break due to subrelativistic expansion (Huang, Dai \& Lu 1998, Ciardi \& Loeb 2000)
can occur relatively early for a dense circumburst medium,
but is irrelevant for the present purpose and not considered here.

\label{lastpage}

\end{document}